\documentclass[a4paper,twocolumn]{article}

\usepackage{graphicx}

\newlength{\figwidth}

\begin{document}

\setlength{\figwidth}{\columnwidth}

\title{Multifractal analysis of the electronic states in the Fibonacci
superlattice under weak electric fields}

\author{M. Wo{\l}oszyn\footnote{woloszyn@agh.edu.pl}, B.J. Spisak}

\date{}

\maketitle

\begin{center}
\emph{AGH University of Science and Technology,\\
Faculty of Physics and Applied Computer Science,\\
al.~A.~Mickiewicza 30, 30-059~Krakow, Poland}       
\end{center}

\abstract{
Influence of the weak electric field on the electronic structure of
the Fibonacci superlattice is considered.
The electric field produces a nonlinear dynamics of the energy
spectrum of the aperiodic superlattice.
Mechanism of the nonlinearity is explained in terms of energy levels
anticrossings.
The multifractal formalism is applied to investigate the effect of weak
electric field on the statistical properties of electronic
eigenfunctions.
It is shown that the applied electric field does not remove the
multifractal character of the electronic eigenfunctions, and that the
singularity spectrum remains non-parabolic, however with a modified shape.
Changes of the distances between energy levels of neighbouring eigenstates
lead to the changes of the inverse participation ratio of the
corresponding eigenfunctions in the weak electric field.
It is demonstrated, that the local minima of the inverse participation
ratio in the vicinity of the anticrossings correspond to discontinuity of the first
derivative of the difference between marginal values of the singularity
strength.
Analysis of the generalized dimension as a function of the
electric field shows that the electric field correlates spatial
fluctuations of the neighbouring electronic eigenfunction amplitudes in
the vicinity of anticrossings, and the nonlinear character of the scaling
exponent confirms multifractality of the corresponding electronic eigenfunctions.
}


\section{Introduction}\label{sec:intro}

Statistical properties of the electronic states in nanosystems are a subject of
great interest
(see, for example, Refs.~\cite{Mirlin_PR326p259,Evers_RMP80p1355,Guhr_PR299p189,Janssen_PR295p1}
and the references therein).
Partially it is due to the progress of experimental methods of nanophysics
which allows to intentionally fabricate high-quality heterostructures consisting
of alternating layers of different materials~\cite{Alferov_RMP73p767,Kroemer_RMP73p783,Milun_RPP65p99,Fert_RMP80p1517}.
The thickness of each layer can be controlled during the growth process
with accuracy of one atomic monolayer, so that one can fabricate
multilayer systems (superlattices) with the desired geometrical
parameters of layers and well defined interfaces.
In this way periodic as well as disordered multilayer systems can be
obtained by the sequential deposition of layers with different thickness of
material.

An intermediate case between periodic and disordered multilayer systems
corresponds to aperiodic order of layers~\cite{Macia_RPP69p397,Vekilov_PU53p537}.
Such structures are intentionally generated by the deposition of
layers of two different materials
according to the Fibonacci, Rudin-Shapiro, Thue-Morse, etc.
sequences~\cite{Luck_JSP53p551,Endo_PRB78p085311,Merlin_PRL55p1768,%
Jarrendahl_PRB51p7621,Chomette_PRL57p1464,%
Laruelle_PRB37p4816,Toet_PRL66p2128,Munzar_JPCM6p4107,Yamaguchi_SSC75p955,%
Lanzillotti_PRB76p174301,Passias_OE17p6636,Taguchi_MRS161p199,Karkut_PRB34p4390}.
Experimental as well as theoretical studies of these superlattices are
concentrated on the consequences of the long-range correlations induced
by the aperiodic arrangement at a length scale longer than atomic
one~\cite{Yuan_PRB62p15569,Rieth_JPCM10p783}.
In particular, this problem has been extensively investigated in the
Fibonacci superlattices which are regarded as a typical example of
aperiodic systems~\cite{Arriaga_JPCM9p8031,Velasco_pssb232p71,Aziz_SSC150p865}.
In these studies, it has been found that the wave functions of
one-particle states are critical, i.e. nor extended neither
localized~\cite{Kroon_PRB66p094204}.
Further studies have shown that the decay of the envelope
wave function obeys the power law and  its structure can be regarded
as a multifractal resembling the case of electronic wave
functions in disordered systems at the mobility
edge~\cite{Mirlin_PR326p259,Evers_RMP80p1355,Castellani_JPA-MG19pL429}.
The shape of the wave function
is highly fragmented in the finite Fibonacci superlattice, and in
the limit of infinite Fibonacci superlattice corresponds to a
self-similar Cantor set with zero Lebesgue
measure~\cite{Kohmoto_PRL50p1870,Kohmoto_PRB35p1020}.

The purpose of this work is to examine the effect of the nonlinear dynamics
of the energy levels driven by weak electric field on the global as well
as local electronic structure of the finite semiconductor Fibonacci
superlattice made of two different semiconductor layers.
Energy spectrum of this superlattice is particularly
interesting in weak electric fields where anticrossings lead to closing of the
energy gaps~\cite{Carpena_PLA231p439,SalazarJPCM22p115501,Spisak_PRB80p035127}.
Particular attention is paid to the effect of this nonlinear dynamics
of energy levels
on spatial fluctuations of the electronic wave function amplitudes and
their spatial extents.

In the systems with broken translational symmetry
the multifractal analysis provides deep insight into the nature
of the electronic wave function~\cite{Martin_PRE82p046206,Mirlin_PRL97p046803,%
Cuevas_PRB68p024206,Cuevas_PRB76p235119,Faez_PRL103p155703,%
Kravtsov_PRB82p161102,FyodorovJSM2009,Pook_ZPB82p295}
and allows to explore
the localization phenomena in the presence of the electric field.

Although the results of the present work are related to the semiconductor
superlattice, they can be also generalized to other aperiodic superlattices,
e.g. photonic or phononic band gap structures under influence of appropriate
perturbations~\cite{HassouaniPRB74p035314,Steurer_JPD40pR229}.

The paper is organized as follows.
In Sec.~\ref{sec:model}, we present the model of the Fibonacci superlattice
and the methods of its analysis,
in Sec.~\ref{sec:results}, we present the results and discussion, and
the conclusions are presented in Sec.~\ref{sec:conclu}.

\section{Model of the Fibonacci superlattice and the methods of analysis}\label{sec:model}

\begin{figure}
\includegraphics[width=\figwidth]{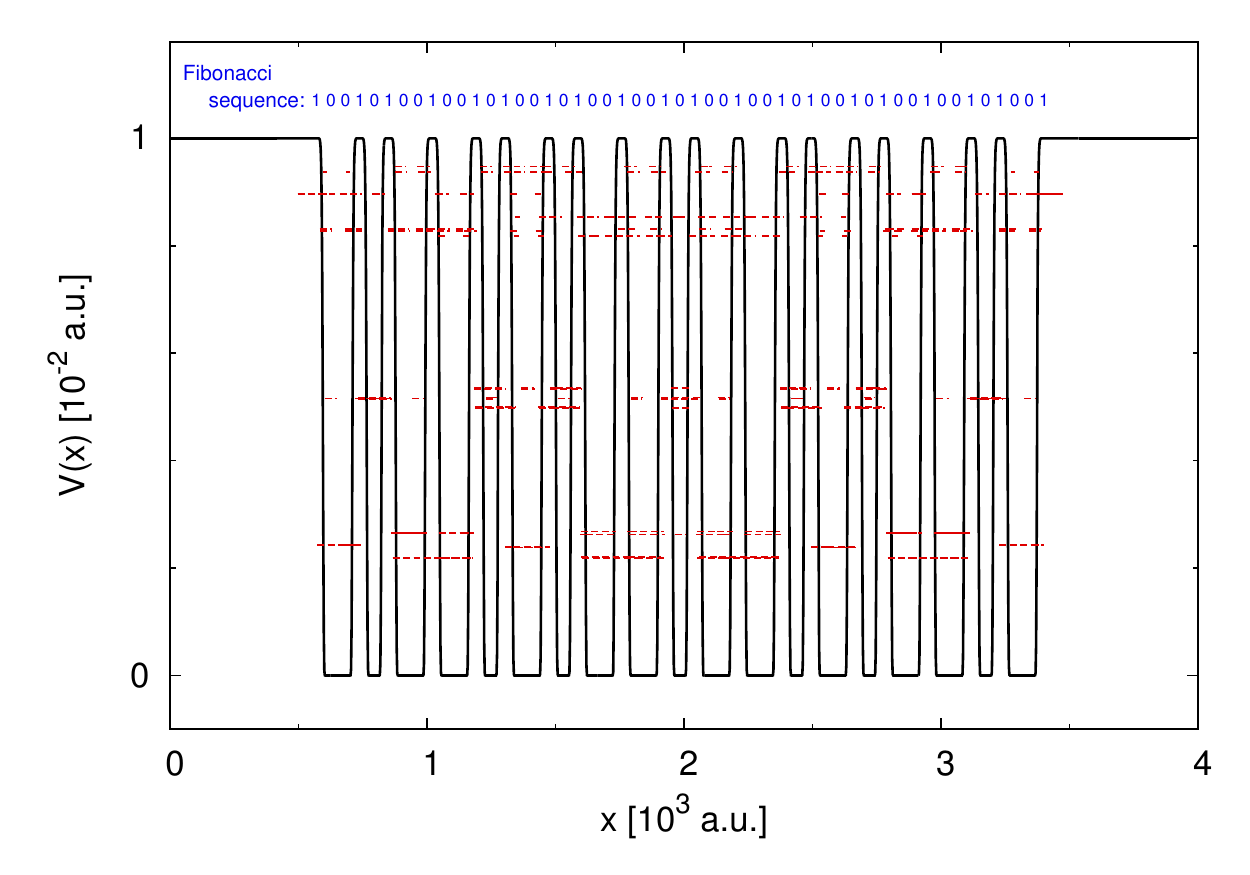}
\caption{\label{fig:potential}(Colour online)
Potential energy profile of the Fibonacci superlattice formed by
20 layers of Al$_{0.3}$Ga$_{0.7}$As and 31 layers of GaAs for
without the external electric field.
The subsequent layers are composed following the Fibonacci
binary sequence, i.e. each Al$_{0.3}$Ga$_{0.7}$As or GaAs layer
corresponds to 1 or 0 in the sequence, respectively.
The electronic states are plotted as (red) horizontal lines at their energies,
with lines plotted for those $x$ at which $|\psi_n^2(x)|$ exceeds its average
value.}
\end{figure}

We consider the semiconductor aperiodic superlattice generated according to the
Fibonacci sequence of two different semiconductor layers made of
Al$_{0.3}$Ga$_{0.7}$As and GaAs~\cite{Merlin_PRL55p1768,Taguchi_MRS161p199}.
The differences in the bandgaps of Al$_{0.3}$Ga$_{0.7}$As and GaAs
semiconductor layers lead to the discontinuities in the
conduction as well as valence band edge profiles at the interfaces~\cite{Sze2007}.
This creates the effective potential energy that consists of the set of barriers
and potential wells distributed along the growth axis of the semiconductor
superlattice (Fig.~\ref{fig:potential}).

The conduction-band potential energy $V(x)$ is
modelled by the superposition of the power-exponential
potentials in the form~\cite{Ciurla_PE15p261}
\begin{equation}
V(x)=\sum_{i=1}^N V_0 \exp{\bigg[-\bigg|\frac{x-x_i}{c} \bigg|^p\bigg]},
\label{eq: p-e potential}
\end{equation}
where $N$ is the number of Al$_{0.3}$Ga$_{0.7}$As barriers located at positions
$x_i$ and having height $V_0$, with parameters $c$ and $p$ characterizing the
shape of barriers.
The positions of barriers, $x_i$, are distributed according to the binary
Fibonacci sequence generated over set $\{0,1\}$ using the following inflation
rules~\cite{Bovier1993,Spisak_APPB38p1951}:
$0\longrightarrow 01$, and $1\longrightarrow 0$.
These rules allow us to obtain the sequence $(0,1,0,0,1,0,1,0,0,1,0,\ldots)$
of any desired length.
In our notation $1$ corresponds to Al$_{0.3}$Ga$_{0.7}$As layer (barrier)
and $0$ corresponds to a single GaAs layer having the same width as the barrier.

\begin{figure}
\includegraphics[width=\figwidth]{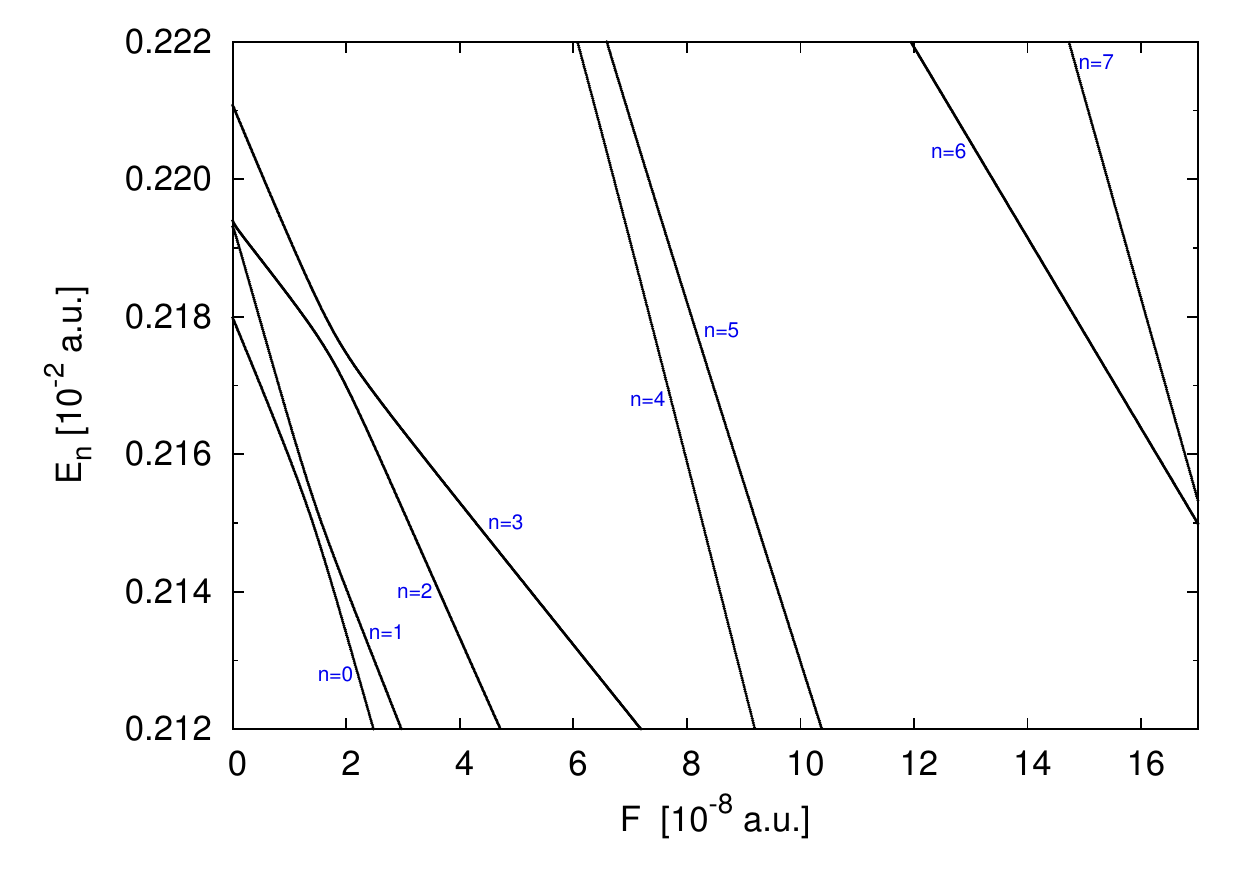}
\caption{\label{fig:levels}(Colour online)
The motion of the eight lowest eigenvalues $(n=0,1,..,7)$ driven by the electric
field in the finite Fibonacci superlattice formed by 20 layers of
Al$_{0.3}$Ga$_{0.7}$As and 31 layers of GaAs.}
\end{figure}

The envelope wave function of one-particle electronic eigenstate of the
Fibonacci superlattice can be found by the solution of the Schr\"odinger
equation within the effective mass approximation
\begin{equation}
-\frac{\hbar^2}{2 m^{\ast}}
\frac{\mathrm{d}^2}{\mathrm{d}x^2}\psi(x)+V(x)\psi(x)=E\psi(x),
\label{eq: Schroedinger}
\end{equation}
where $m^{\ast}$ is the effective mass of the conduction band electron.

The Schr\"odinger equation~(\ref{eq: Schroedinger}) with the potential energy
$V(x$) given by equation~(\ref{eq: p-e potential}) and the Dirichlet boundary
conditions, $\psi(0)=\psi(L)=0$ (where $L$ is the size of the computational
box), constitutes the eigenvalue problem which can be solved numerically.
This form of the boundary conditions allows us to neglect the complexity of the
energy spectrum due to the surface states.

The energy spectrum of the Fibonacci superlattice formed from a finite
number of  barriers and quantum wells is characterized by the
density of states which can be expressed as follows
\begin{equation}
\mathrm{DOS}(E)=\sum_{n}\delta(E-E_n).
\label{eq: DOS}
\end{equation}
This quantity gives only global properties of the system, whereas the local
properties can be described by the local density of states which directly
explores the local amplitude of the electronic wave function for a given
energy.
The local density of states (LDOS) is defined by the formula
\begin{equation}
\mathrm{LDOS}(x,E)=\sum_{n}|\psi_n(x)|^2\delta(E-E_n),
\label{eq: LDOS}
\end{equation}
where
$\psi_n(x)$ is the normalized electronic
wave function of the $n$-th eigenstate.

One of the most characteristic properties of the electronic states in the finite
Fibonacci superlattice are the self-similar spatial fluctuations of the electronic
wave function amplitude which stem from the aperiodic distribution of the
barriers and wells in the system.
These fluctuations can be analyzed by the multifractal formalism which allows
one a deeper insight into their nature.
An essential ingredient of the multifractal formalism is the normalized
probability measure of the wave function in the $k$-th box $B_k$ of linear size
$\varepsilon$,
\begin{equation}
P_{nk}(\varepsilon) = \int_{B_k(\varepsilon)} \mathrm{d}x \, |\psi_n(x)|^2.
\label{eq: box}
\end{equation}
The multifractal analysis of the electronic wave functions in the Fibonacci
superlattice can be performed efficiently by applying the box-counting
procedure~\cite{Chhabra_PRL62p1327,Falconer2003}
with condition: $a \ll \varepsilon \ll L$ ($a$ is the lattice
constant).
The normalized $q$-th moment of the probability measure of the wave function is
given by the formula
\begin{equation}
\mu_{nk}(\varepsilon; q)=\frac{P_{nk}^q(\varepsilon)}
{\sum_{j=1}^{N(\varepsilon)}P_{nj}^q(\varepsilon)},
\label{eq: q-th moment}
\end{equation}
where $N(\varepsilon)=L/\varepsilon$ is the number of boxes.

For each value of the scaling index $q$ we can determine the singularity
strength according to the formula
\begin{equation}
\alpha_{n}(q)=\lim_{\delta\rightarrow 0}
\frac{
\sum_{k=1}^{N(\varepsilon)}
\mu_{nk}(\varepsilon; q)\ln{\mu_{nk}(\varepsilon; 1)}
}{\ln{\delta}}
\label{eq: alpha-strength}
\end{equation}
and the corresponding singularity spectrum in a parametric representation as
follows
\begin{equation}
f(\alpha_n)=\lim_{\delta\rightarrow 0}
\frac{
\sum_{k=1}^{N(\varepsilon)}
\mu_{nk}(\varepsilon; q)\ln{\mu_{nk}(\varepsilon; q)}
}{\ln{\delta}},
\label{eq: f-spectrum}
\end{equation}
where $\delta=\varepsilon/L$ is the ratio of the box size to the system
size.

The singularity spectrum $f(\alpha_n)$ gives an accurate description of the
multifractal properties of the probability measure of the wave function.
Instead of the singularity spectrum, we can equivalently
consider a hierarchy of generalized dimensions of the wave function.
It stems from the fact that the generalized dimension $D_n(q)$ is related to the
singularity spectrum by the formula
\begin{equation}
D_n(q)=
\frac{f(\alpha_n)-q\alpha_{n}(q)}{1-q}.
\label{eq: D-dimension}
\end{equation}
For the integer values of the scaling index $q$, the generalized dimensions of
the wave function have a physical meaning~\cite{Hentschel_P8D435}.
A particularly interesting generalized dimension corresponds to $q=2$ when
it is known as the correlation dimension of the wave function.
Its significance results from the relation to the density-density correlation
function and the inverse participation ratio (IPR) which is a measure of the
spatial extent of the eigenstate and can be used to describe the
localization properties of the electronic eigenstates in real
space~\cite{Huckestein_RMP67p357}.
It is an important point since the localization length cannot be defined through
the exponential spatial decay of the electronic wave function in the finite
aperiodic or disordered superlattices.
This fact is a consequence of the strong spatial fluctuations of electronic wave
function amplitude and therefore a more adequate description is based on the
spatial extension of the wave functions given by the
IPR parameter~\cite{Spisak_APPB38p1951,Wegner_ZPB36p209,Thouless_PR13p93},
\begin{equation}
\mathrm{IPR}_n=\int_0^L dx \, |\psi_n(x)|^4.
\label{eq: IPR}
\end{equation}

A combination of equations~(\ref{eq: alpha-strength}), (\ref{eq: f-spectrum})
and (\ref{eq: D-dimension}) leads to the relation between
the generalized dimension of the electronic wave function and the
scaling exponent, $\tau_n(q)$, for the $q$-th moment of the probability
measure~\cite{Pook_ZPB82p295}, namely
\begin{equation}
\tau_n(q)=D_n(q)(1-q).
\label{eq: scaling exponents}
\end{equation}
Deviation of the scaling exponent from the linear function of $q$
signals the multifractality of the electronic state.

In the limit of weak electric field, the interband transitions can be neglected
and the influence of the electric field on the electronic states of
Fibonacci superlattice can be considered by including an additional perturbation
in the form
\begin{equation}
W=-eFx \, ,
\label{eq: perturbation}
\end{equation}
where $e$ is the elementary charge, and $F$ is the external homogeneous electric
field applied along the growth axis of the system.

In fact, the weak electric field is treated non-perturbatively as a
parameter which produces the motion of the energy levels as it is shown
in Figure~\ref{fig:levels}.
For the considered superlattice the level repulsion leads to the formation
of anticrossings in the energy spectrum, which is
a simple consequence of the dimensionality of the system and the
Dirichlet boundary conditions applied to the Schr\"odinger equation.

Introducing the perturbation to the Schr\"odinger equation
(\ref{eq: Schroedinger}) finally leads to the equation
\begin{equation}
-\frac{\hbar^2}{2 m^{\ast}}
\frac{\mathrm{d}^2}{\mathrm{d}x^2}\psi(x)+[V(x)-eFx]\psi(x)=E\psi(x),
\label{eq: SchroedingerWithField}
\end{equation}
which allows us to investigate the effect of weak electric field
on the global (energy spectrum, DOS) as well as local properties
(LDOS, spatial fluctuations of the wave function amplitude)
of the finite Fibonacci superlattice.

\section{Numerical results and discussion}\label{sec:results}

Using the model of the finite Fibonacci superlattice described in
Sec.~\ref{sec:model} we have investigated its global as well as local
electronic properties in the limit of weak electric field.
All numerical values of the physical quantities which are considered here are given
in the atomic units, i.e., $\hbar=|e|=m_0=1$.
The parameters of the model potential defined in equation~(\ref{eq: p-e potential})
are taken to be $V_0=0.27$~eV = $0.01$~a.u.,  $c=1.5$~nm = $28.3$~a.u.,
and $p=10$, which means that the system
consists of Al$_{0.3}$Ga$_{0.7}$As barriers having width of 3~nm and
separated by GaAs layers of 3~nm or 6~nm width (the latter in case of two
consecutive zeros in the Fibonacci binary sequence).
The number of barriers is $N=100$ and the constant effective mass approximation
is used with the value of $m^{*}$ equal $0.067 m_0$.

\begin{figure}
\includegraphics[width=\figwidth]{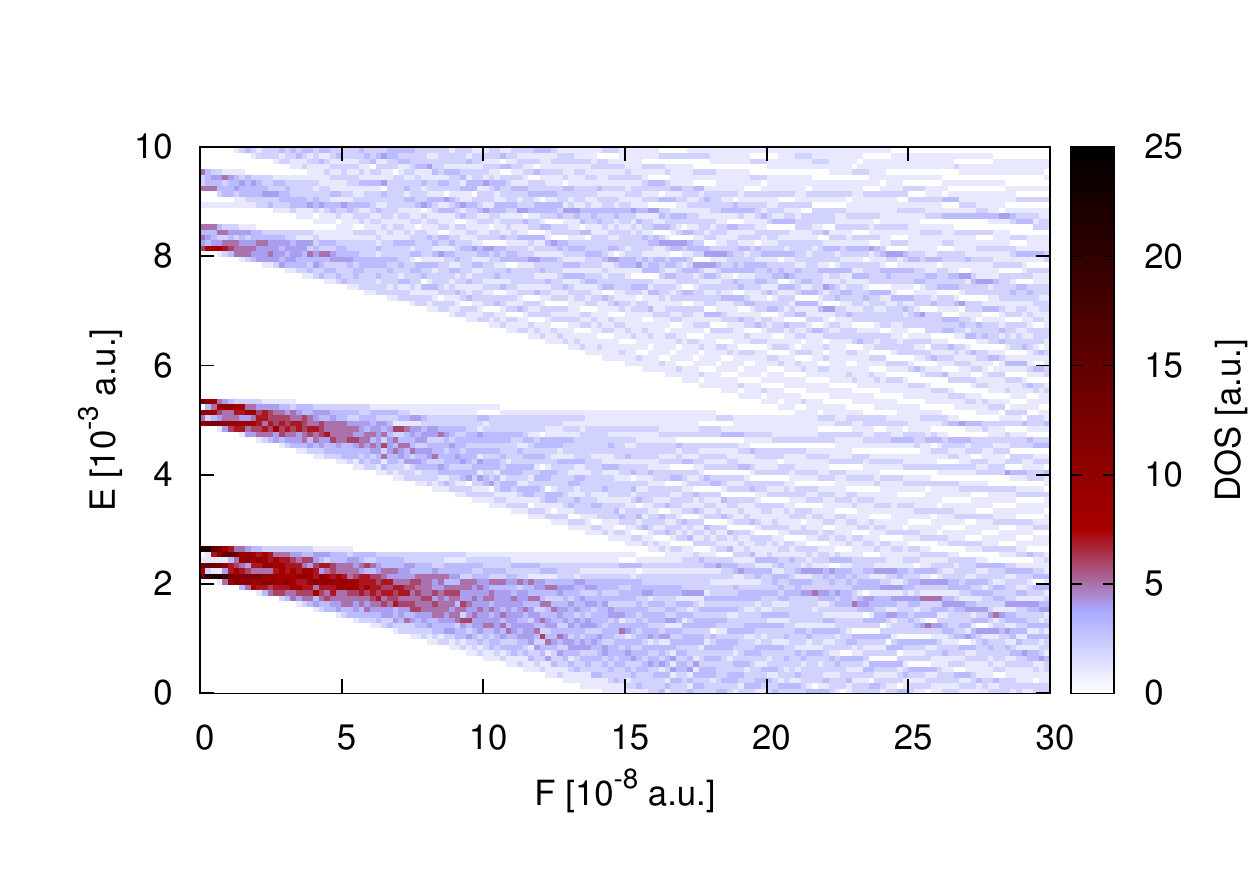}
\caption{\label{fig:dos}(Colour online)
Density of states (DOS) as a function of energy $E$ and electric field $F$.
The results for the Fibonacci superlattice with $N=100$.}
\end{figure}

\begin{figure}
\includegraphics[width=\figwidth]{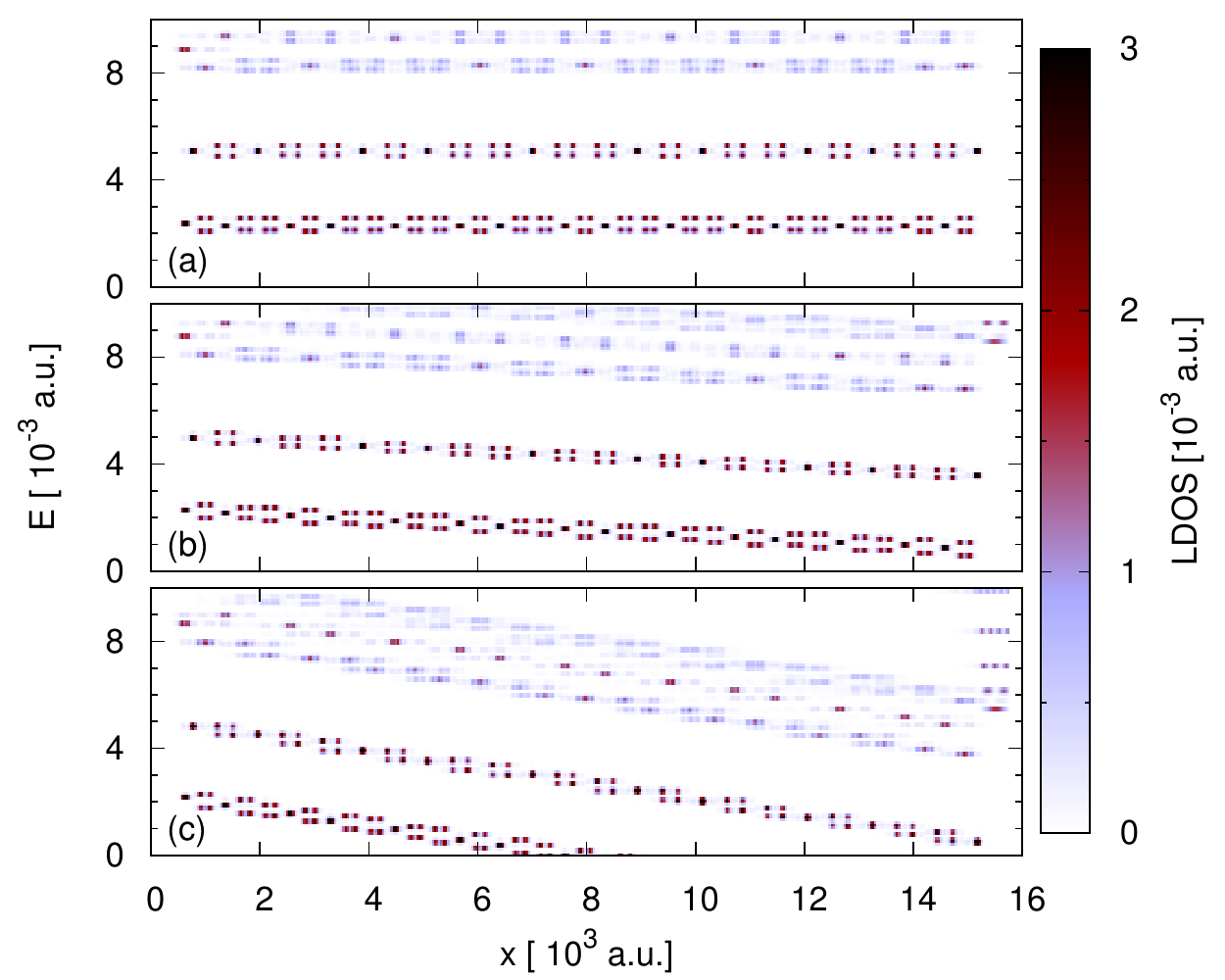}
\caption{\label{fig:ldos}(Colour online)
Local density of states LDOS for the Fibonacci superlattice with $N=100$
as a function of energy $E$ and position $x$.
(a)~$F=0$, (b)~$F=10^{-7}$~[a.u.] and (c)~$F=3 \times 10^{-7}$~[a.u.].}
\end{figure}

One of the characteristic features of the obtained energy spectrum is the
presence of minibands with fractal structure (Fig.~\ref{fig:dos}).
When the electric field is increased these minibands become broader, with
a nonuniform distribution of the energy levels occurring during the entire process.
It is the reason why darker and lighter areas corresponding to dense
and rare subsets of the levels are present within the bands
in Figure~\ref{fig:dos} showing the influence of the electric field on the DOS
of the Fibonacci superlattice.
On the other hand, the general structure of the LDOS is not altered and
the effect of the electric field is restricted to the change of the slope along the
horizontal axis (see Figure~\ref{fig:ldos} presenting how the electric field modifies
LDOS, which leads to the previously mentioned broadening of the minibands).

\begin{figure}
\includegraphics[width=\figwidth]{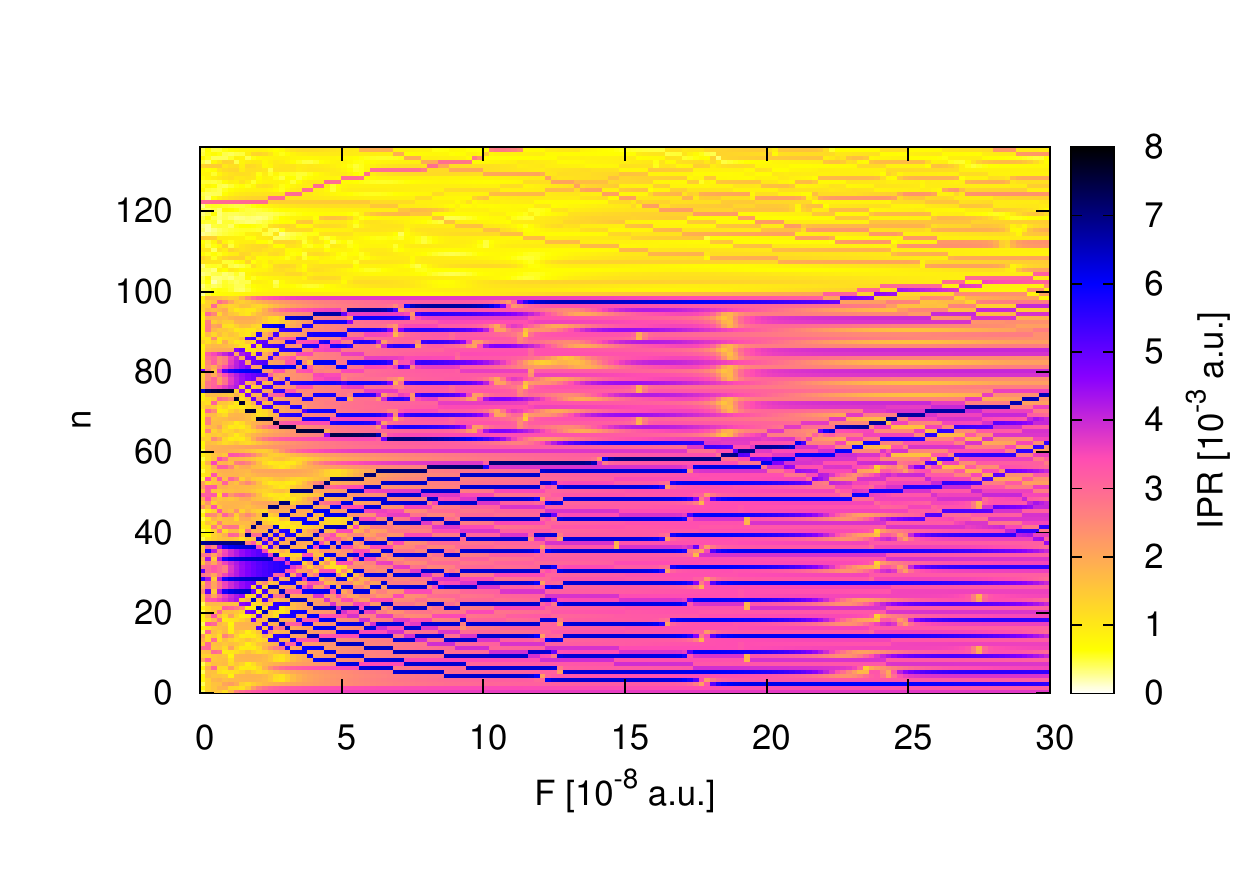}
\caption{\label{fig:ipr}(Colour online)
Inverse participation ratio $\mathrm{IPR}$ calculated for $n=0,\ldots,136$
(i.e. for all eigenstates $\psi_n$ which are bound for $F=0$)
in case of the Fibonacci superlattice with $N=100$.}
\end{figure}

The presence of the anticrossings between the energy levels is the direct cause of the
nonuniform structure of the energy spectrum as well as the density of states
function~\cite{Spisak_PRB80p035127}.
In the region of anticrossings, the electronic wave functions change the degree
of localization measured by the IPR parameter.
Figure~\ref{fig:ipr} shows the IPR parameter for
electronic eigenstates of the Fibonacci superlattice as a
function of electric field.
The lowest degree of localization is observed in the absence of the electric
field, but in contrast to finite periodic systems the localization degree is not simply
increasing with the electric field.
In case of the finite Fibonacci superlattice the observed behaviour is much more complex.
Instead of the successive states with very similar IPR dependence on the
electric field, we can notice that for a chosen value of the electric field
the highly localized states are separated by the states with much weaker
localization.

Therefore, the electric field affects the degree of localization, which
in turn results in the change of the spatial fluctuations of the
wave function.
In order to explain the relation between the increasing electric field and the
modification of the spatial fluctuations of the wave functions we have performed
a detailed analysis of the lowest energy levels, i.e. the ground-state level and
the first two excited state levels.
It should be noted, that each level is changed not only by the external electric
field, but also by the coupling to the nearest energy levels.
As a result, a nonlinear dependence of the energy levels on the
electric field is observed.

\begin{figure}
\includegraphics[width=\figwidth]{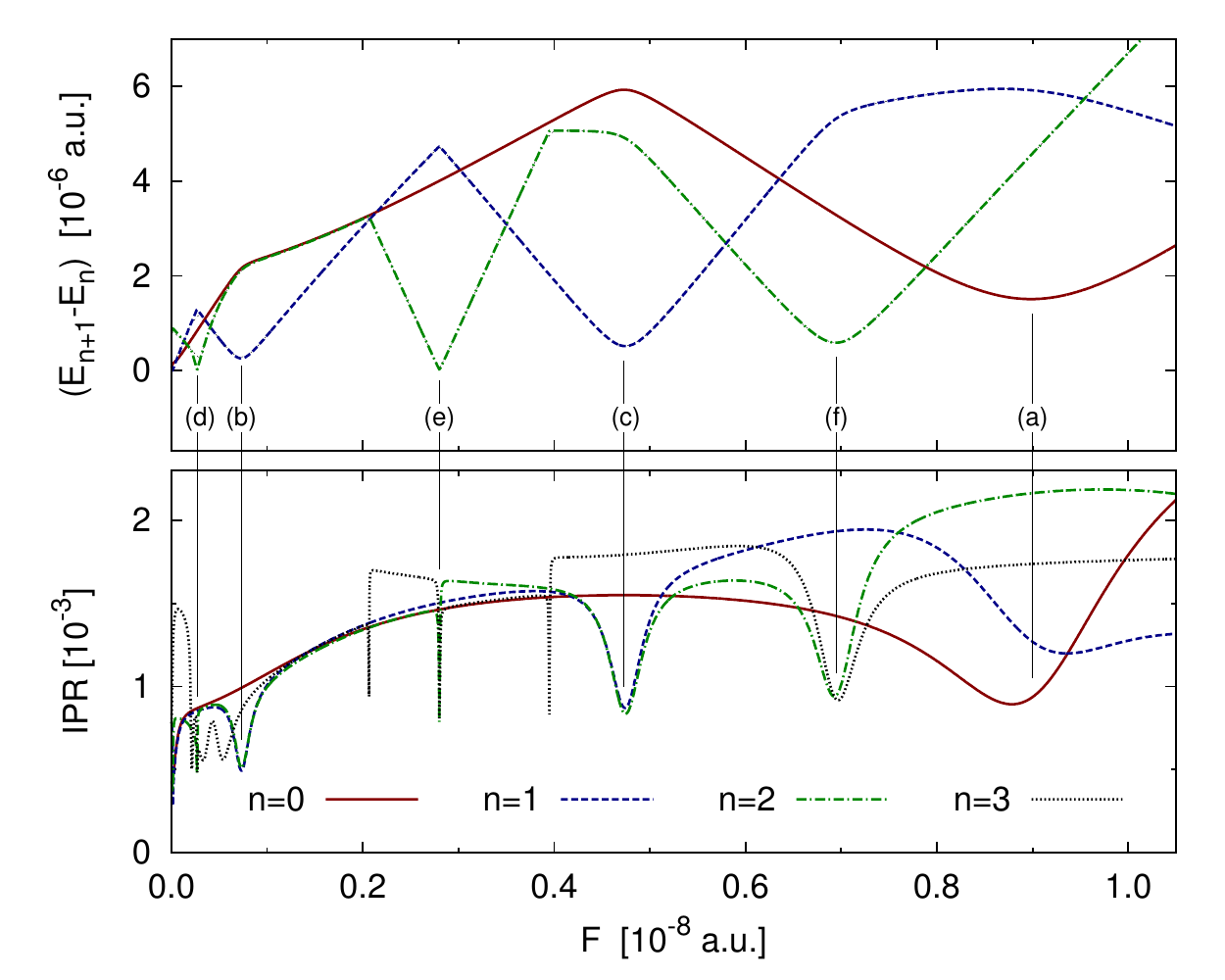}
\caption{\label{fig:delta}(Colour online)
Differences $(E_{n+1}-E_n)$
between the energies of the four lowest subsequent levels (upper panel) and
the corresponding values of the inverse participation ratio IPR (lower panel)
in the case of $N=100$ Fibonacci superlattice.
Letters (a)-(f) denote the positions of anticrossings.
}
\end{figure}

The distances between the energy levels $n$ and
$n+1$, calculated for  $n=0,1,2$, are presented in Figure~\ref{fig:delta}.
The neighbouring states tend to change their positions in such a way that the
distances between them increase or decrease alternately.
This kind of analysis may be also generalized for higher energy levels,
however, in this case a large number of anticrossings is observed which
makes the analysis not particularly suitable for the purpose of the clear
explanation of the phenomenon.
For this reason the further discussion focuses on the regions
of anticrossings labelled by (a)-(f) in Figure~\ref{fig:delta}.

Figure~\ref{fig:delta} presents also the values of IPR parameter calculated
for the corresponding states in the same range of the electric field.
For the values of the electric field at which the anticrossings are observed,
i.e. values of $(E_{n+1}-E_n)$ have minima, also the values of IPR parameter
for $n$ and $n+1$ have local minima.
It means that the electronic wave functions matching those levels become
less localized and occupy a larger region of space due the coupling between
the closely lying levels.

The further study of the properties of electronic wave functions corresponding
to the chosen states is based on the multifractal analysis.
The values of the singularity strength $\alpha_n$ and the corresponding
singularity spectrum $f(\alpha_n)$, parametrized by the electric field,
defined in equations~(\ref{eq: alpha-strength}) and (\ref{eq: f-spectrum}), are
calculated using the standard 'box-counting' procedure repeated for the
values of the electric field changing from $0$ up to $10^{-8}$a.u.

\begin{figure}
\includegraphics[width=\figwidth]{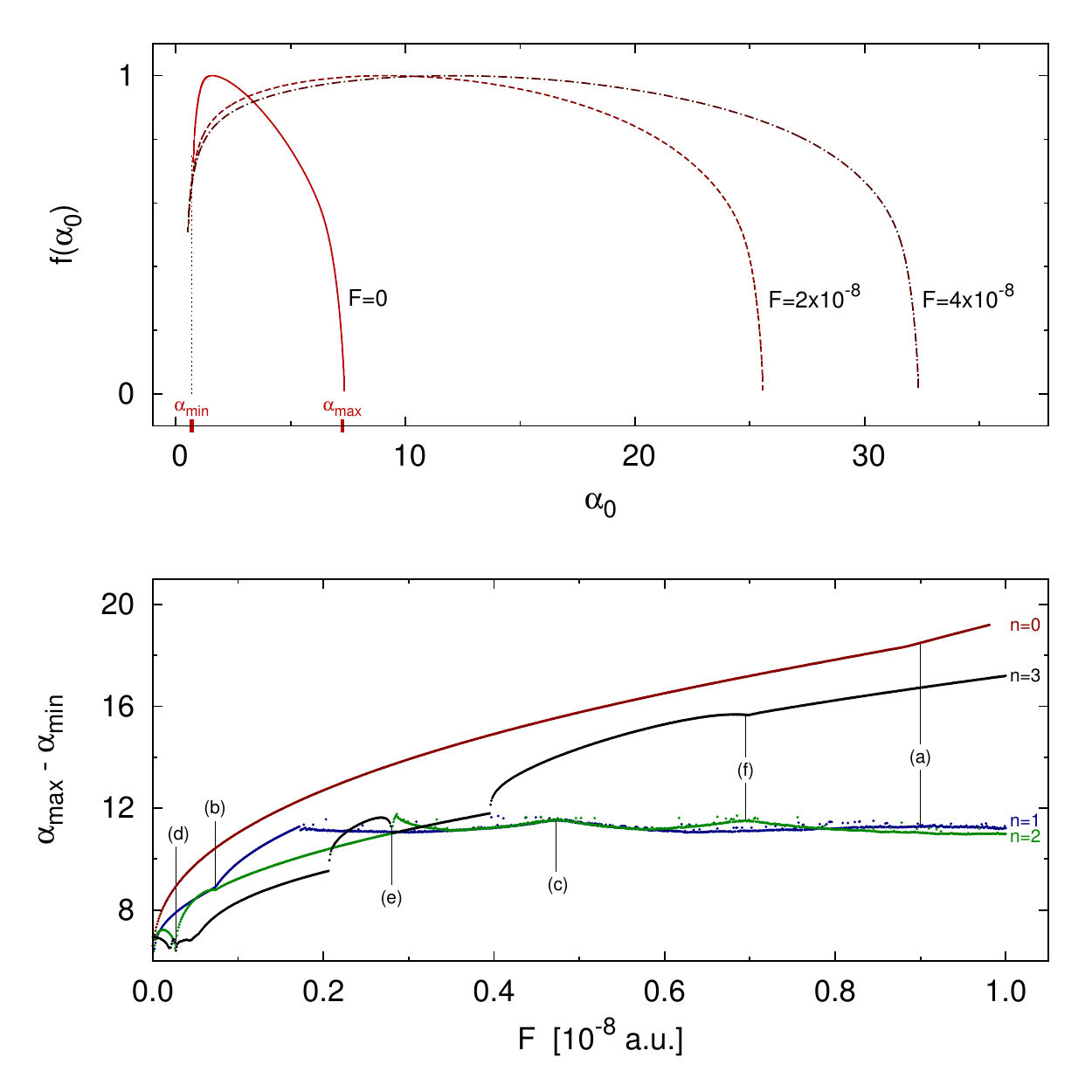}
\caption{\label{fig:falpha}(Colour online)
Upper panel:
Multifractal spectrum $f(\alpha_n)$ calculated for the ground level state $n=0$
at different values of the electric field $F$ with $\alpha_{\mathrm{min}}$ and
$\alpha_{\mathrm{max}}$ indicated for $F=0$.
Lower panel:
The difference between the maximal and minimal singularity strengths,
$\alpha_{\mathrm{max}}-\alpha_{\mathrm{min}}$, for the first four lowest energy
levels, $n=0,1,2,3$, with letters (a)-(f) denoting the positions of
anticrossings (see also Figure~\ref{fig:delta}).
}
\end{figure}

The $f(\alpha_n)$ spectrum for the Fibonacci superlattice is found to be
strongly asymmetric when no electric field is applied, as it is shown for $n=0$
in Figure~\ref{fig:falpha}.
However, while $f(\alpha_n)$ is broadened when the electric field is present,
the minimal value of the singularity strength, $\alpha_{\mathrm{min}}$,
remains almost constant and close to zero, which is its minimal possible value.
This fact may be connected with the nature of the discussed eigenfuctions,
which have the localized peaks almost not affected by the increasing electric
field, as it can be seen in Fig.~\ref{fig:ldos}.
Similar behaviour is observed also for higher eigenstates (see the lower
panel of the Figure~\ref{fig:falpha}), and resembles freezing of the
$\alpha_{\mathrm{min}}$ value occurring in disordered
systems~\cite{FyodorovPA2010}.

As a result, the value $(\alpha_{\mathrm{max}}-\alpha_{\mathrm{min}})$
is not an objective measure of the electronic wave function
localization~\cite{Biswas_PLA262p464} in the case of this type of Fibonacci
sequence-based systems under influence
of the electric field, which is clearly visible from the
comparison with the values of the IPR parameter calculated for the
same states (Fig.~\ref{fig:delta}).
Moreover, the first derivative of
$(\alpha_{\mathrm{max}}-\alpha_{\mathrm{min}})$ is not continuous at
the values of the electric field where the anticrossings between the
energy levels are observed.
In light of these results, we rather think that for this kind of systems, the value
$(\alpha_{\mathrm{max}}-\alpha_{\mathrm{min}})$  can be used to detect
the local minima of  the IPR parameter or equivalently to detect
the anticrossings in the energy spectrum.

\begin{figure}
\includegraphics[width=\figwidth]{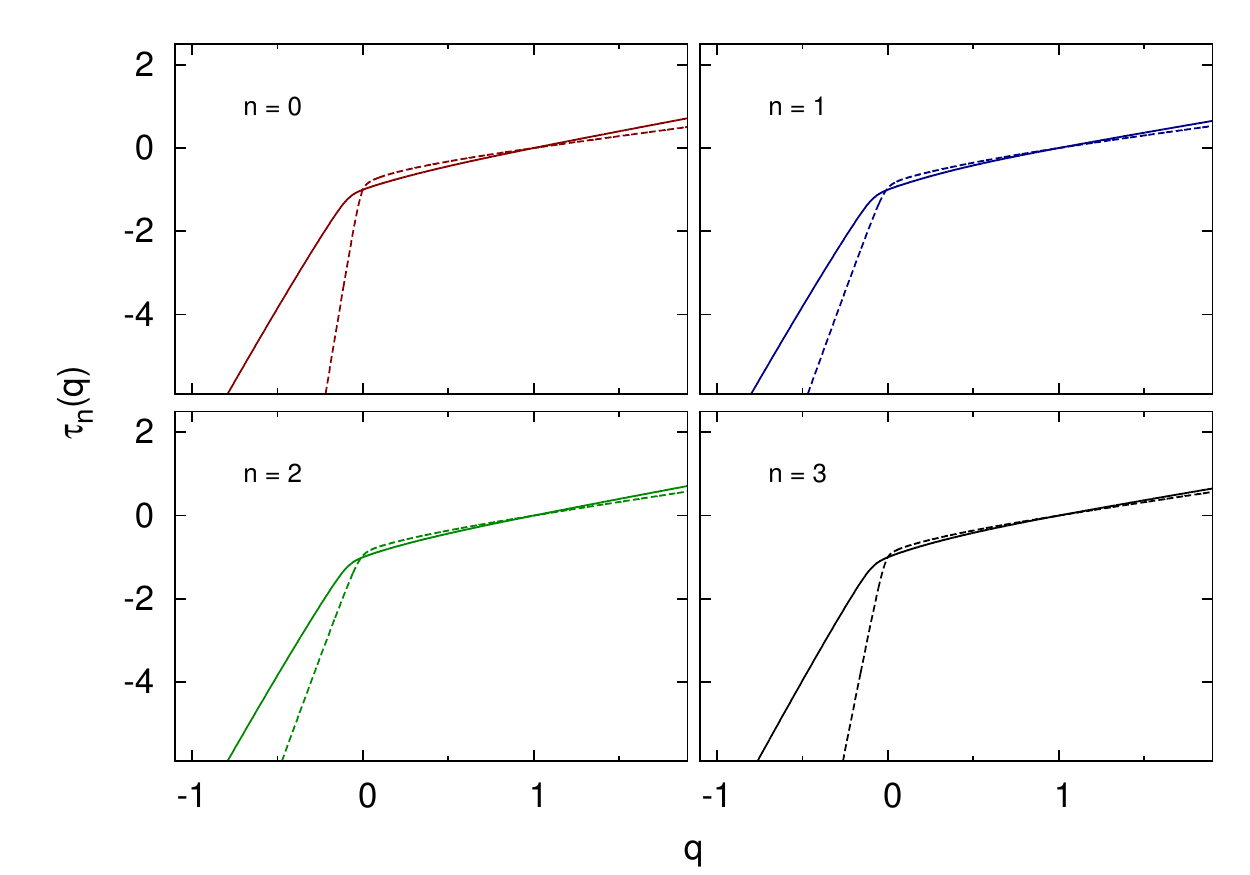}
\caption{\label{fig:tau}(Colour online)
The scaling exponent $\tau_n(q)$ for the electric field $F=0$ (solid line) and
$F=2\times 10^{-8}$ (dashed line) and for states $n=0,1,2,3$.
}
\end{figure}

The multifractal character of the electronic wave function in the
finite Fibonacci superlattice and the influence of the weak
electric field  becomes much more visible in Figure~\ref{fig:tau}, where the
$q$ dependence of the scaling exponent, $\tau_n(q)$, for a few lowest electronic
states is presented.
In all cases the scaling exponent is a convex function, monotonically increasing
with $q$.
The electric field modifies the slopes of the considered scaling exponents:
to a larger extent for negative $q$, whereas for positive values of $q$ the
changes of the slope are much smaller.
This type of changes of the slopes is related to
marginal values of the singularity strength~\cite{Evers_RMP80p1355},
$\alpha_{\mathrm{min}}$ (being almost constant as a function of the electric field)
and $\alpha_{\mathrm{max}}$ (increasing notably in the electric field).

\begin{figure}
\includegraphics[width=\figwidth]{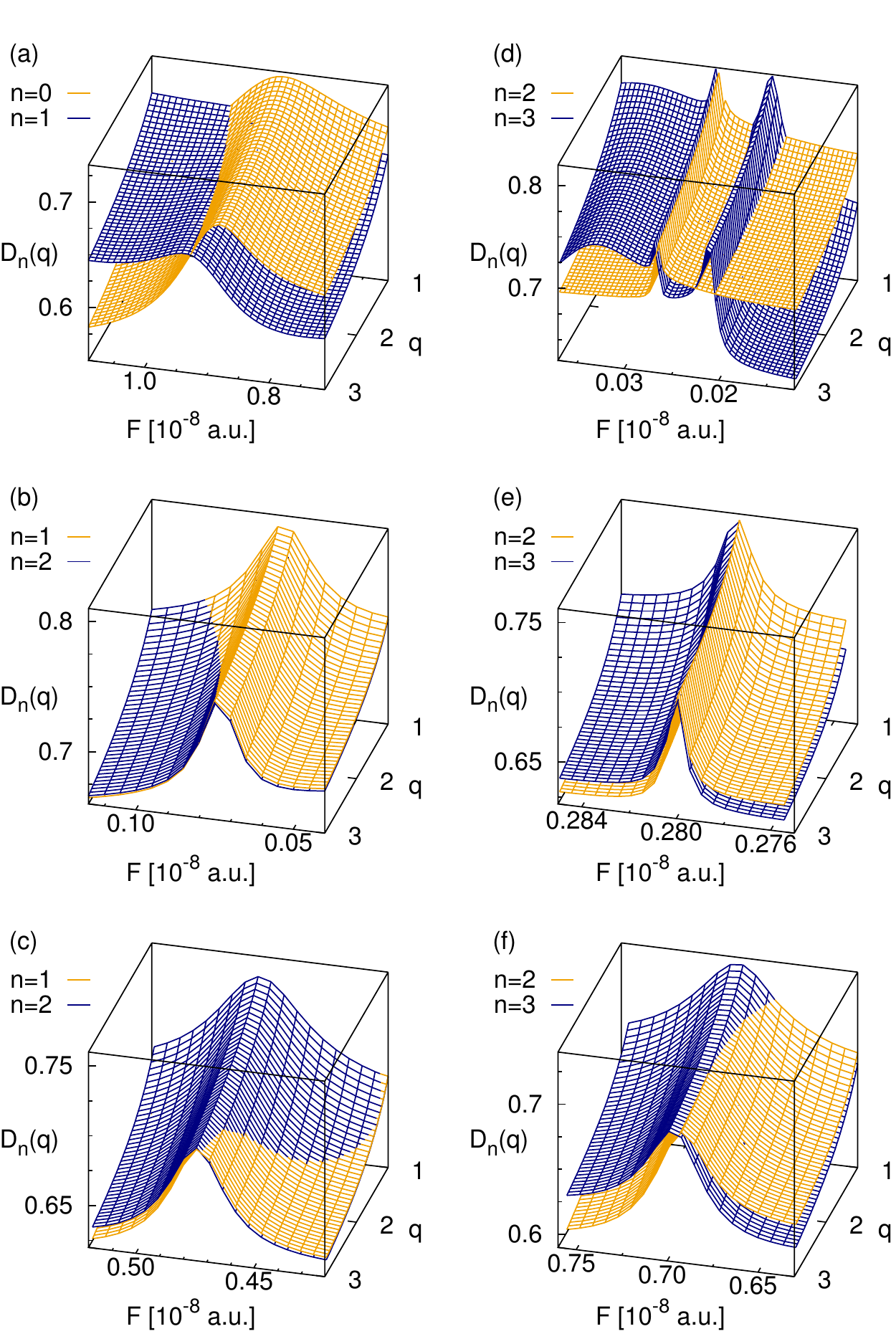}
\caption{\label{fig:Dq}(Colour online)
Generalized dimension $D_n(q)$ calculated in the vicinities of the anticrossings
(a-f) shown in Figure~\ref{fig:delta}.
}
\end{figure}

Changes of $D_n(q)$ in the electric field are also noteworthy, as it is shown
in Figure~\ref{fig:Dq}.
In general, $D_n(q)$ decreases nonlinearly with $q$ and reaches a
constant for $q \to -\infty$, however the value strongly depends on the
state number on the other hand.
For $q \to +\infty$, $D_n(q)$ tends to a constant value, but the dependence of
this value on the electronic state is rather weak.
The nonlinear decrease of $D_n(q)$ with increasing $q$ is a presage of
the wave function multifractality~\cite{Pook_ZPB82p295}.

In the following discussion we present results obtained for $1<q<3$
which is the range where the fluctuations are mostly pronounced.
Figure~\ref{fig:Dq} shows the generalized dimension calculated in
the vicinity of the anticrossings marked in Figure~\ref{fig:delta}.
$D_n(q)$ decreases with $q$, but it has maxima
when analyzed as a function of the electric field.
The maximal values of $D_n(q)$ appear at the same values of the electric field
as the anticrossings and are accompanied by the minima of IPR parameter.
It leads to the conclusion that the maximum of the generalized
dimension $D_n(q)$ as a function of the electric field
corresponds to the decrease of the wave function
localization degree.
Moreover, if we analyze the surfaces of $D_n(q)$ for both states
involved in the anticrossing and plot them as functions of $q$ and the
electric field $F$, it turns out that they cross and this crossing
does not take place at the same value of the electric field for all
values of $q$.
The curve defined by the crossings that joins points for which the generalized
dimensions are the same for the two discussed states, corresponds
to the equality of the generalized moments $\mu_{nk}$
of the probability densities of the electronic states which are
equivalent to the intensity fluctuations~\cite{Prigodin_PRL80p1944}.

\section{Concluding Remarks}\label{sec:conclu}

In the present paper, we have investigated the influence of the weak
homogeneous electric field on the one-electron states in the
finite superlattice generated according to the Fibonacci sequence
of two types of semiconductor layers.
For clarity of presented analysis only a few lowest-energy eigenstates
have been considered in details, but conclusions can be generalized
for higher-energy eigenstates.

We have shown that the nonlinear character of the energy levels dynamics
results from the large number of anticrossings.
Therefore the energy spectrum of the Fibonacci superlattice and the density of
states are nonuniform in the limit of weak electric field.
In the vicinity of the anticrossings,
the inverse participation ratio of the electronic
wave functions for individual electronic eigenstates possesses minima.
These minima of the inverse participation ratio are related to the
maxima of generalized dimension calculated as a function of electric field,
that are observed for $0<q<3$ in the vicinity of the anticrossings.
Moreover, we have shown that the change of the
spatial extent of the electron wave function in the vicinity of the anticrossings
is preceded by the correlation between the spatial fluctuations of
wave functions corresponding to the electronic eigenstates participating in
the anticrossings.

The relation between the positions of anticrossings and the properties of
the singularity spectrum has been revealed, and the difference between the
maximal and minimal values of the singularity strength has been found not to be
a proper measure of the spatial extents of electronic
wave functions in the presented case.
Quite surprisingly, we have found the discontinuity of the first
derivative of the difference between marginal values of the singularity
strength for the values of the electric field where the local minima
of inverse participation ratio exist.
We have also shown that a strong asymmetry of the singularity spectrum is
preserved for all considered values of the electric field,
although the electric field modifies the shape of the spectrum.
This analysis of the relations between the singularity spectrum
and the scaling exponents, together
with the calculated generalized dimensions,
have allowed us to show that the multifractal character of the electronic
wave functions in this type of Fibonacci sequence-based systems
is not destroyed by the weak electric field.

The results presented above correspond mainly to one arbitrarily chosen length
of the Fibonacci sequence used as a basis for the model potential.
Our further numerical studies performed for several different sizes of the
systems show that the same characteristic features are observed, and thus
we have decided to choose a typical example allowing the detailed analysis.

We hope that the presented results can be useful
in a deeper understanding of the nonlinear properties of the energy spectrum
in the aperiodic photonic, phononic or semiconductor superlattices
under influence of an appropriate external perturbation
and motivate the experimental verification of the results.


\section*{Acknowledgement}
Supported by the Polish Ministry of Science and Higher Education and its grants
for Scientific Research.




\end{document}